\documentclass[showpacs,prl,twocolumn,floatfix]{revtex4}
\usepackage{graphicx, amsmath, amssymb, subfigure}

\begin{document}

\title{Transport on coupled spatial networks}

\author{R.~G.~Morris and M.~Barthelemy}

\affiliation{Institut de Physique Th\'{e}orique, CEA, CNRS-URA 2306, F-91191, 
Gif-sur-Yvette, France}

\begin{abstract}
  Transport processes on spatial networks are representative of a
  broad class of real world systems which, rather than being
  independent, are typically interdependent.  We propose a measure of
  utility to capture key features that arise when such systems are
  coupled together.  The coupling is defined in a way that is not
  solely topological, relying on both the distribution of sources and
  sinks, and the method of route assignment.  Using a toy model, we
  explore relevant cases by simulation.  For certain parameter values,
  a picture emerges of two regimes.  The first occurs when the flows go
  from many sources to a small number of sinks.  In this case, network
  utility is largest when the coupling is at its maximum and the
  average shortest path is minimized.  The second regime arises when
  many sources correspond to many sinks.  Here, the optimal coupling
  no longer corresponds to the minimum average shortest path, as the
  congestion of traffic must also be taken into account.  More
  generally, results indicate that coupled spatial systems can give
  rise to behavior that relies subtly on the interplay between the
  coupling and randomness in the source-sink distribution.
\end{abstract}


\maketitle

Systems that can be represented as a group of interacting networks are
found everywhere in modern life~\cite{MH01,SMR+04,VRLI+08}.  From
so-called smart power grids---which couple electrical distribution
networks with ICT control networks~\cite{VV+10}---to interactions
between other types of critical infrastructure networks, such as food
and water supply, transport, fuel, and financial transactions. Recent
theoretical studies on the subject have generated a great deal of
interest by demonstrating that coupling two or more networks together
can lead to system-wide behaviour which differs fundamentally from the
behaviour of each individual 
network~\cite{RP+10Jul,SVB+10Apr,XHJG+11,CG+11Aug,JGSB+11,CDB+12Feb,CDB+12Apr,Boguna:2012, 
resp1,resp2}.  These studies describe essentially cascade-like processes where 
typically either inverse-percolation~\cite{SVB+10Apr,JGSB+11,CDB+12Apr} or 
sandpile
methods~\cite{CDB+12Feb} are used (or variants thereof).  In the
former case, the robustness of the system is characterized by size of
the remaining giant connected component, whilst for the latter, it is
the size of the largest sand-cascade.  In both cases, the quantity of
interest is directly related to the topology of the network and does
not permit any consideration of dynamical processes which may take
place on the network. Furthermore, robustness against cascade failures
is not the only consideration for those affected by such real world
systems.

\begin{figure}[t] \centering \includegraphics[scale = 0.35]{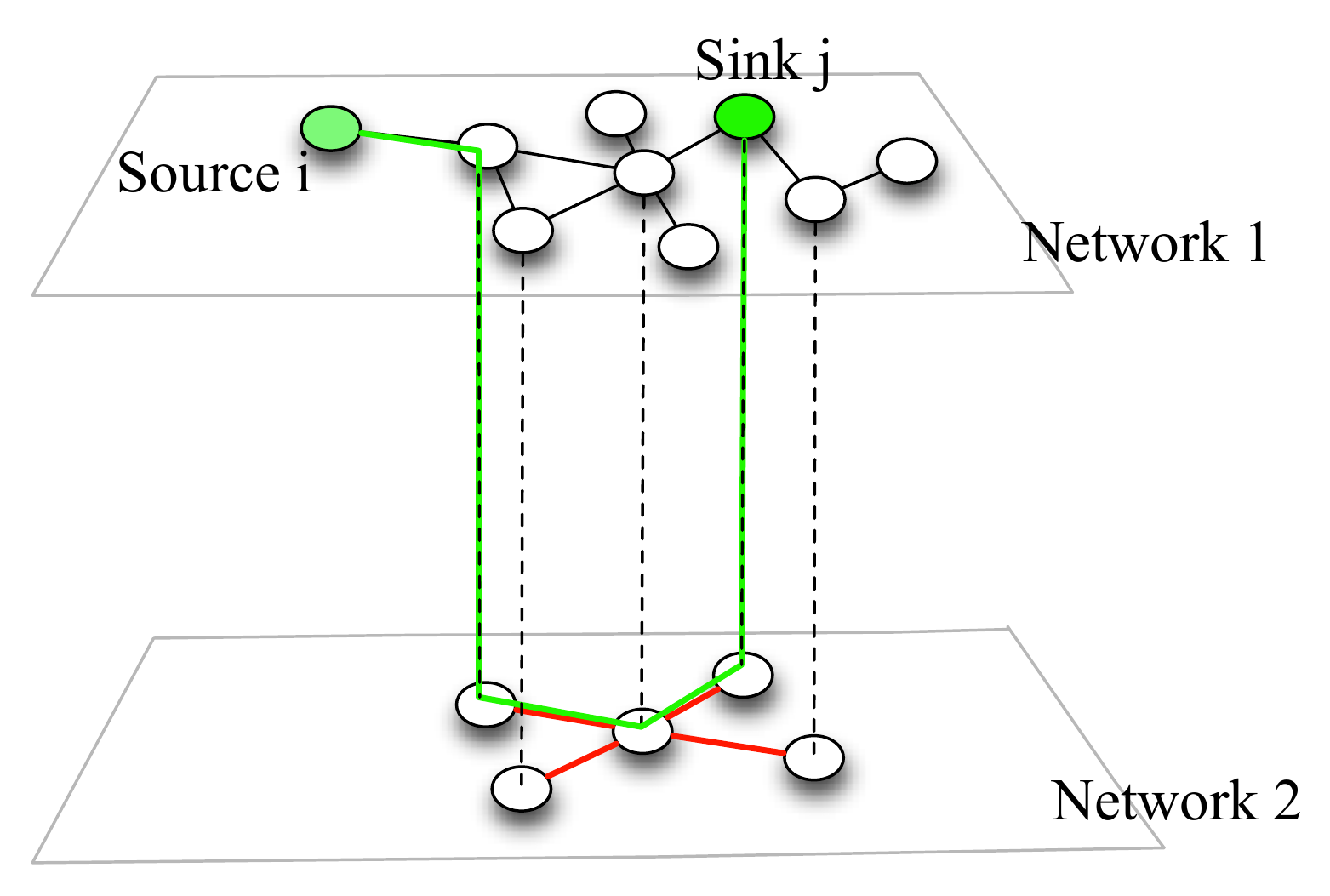}
  \caption{(Color online) A system made of two coupled networks where
    the nodes of network 2 form a subset of the nodes of network 1.
    Edges of network 1 are shown in black, edges of network 2 are
	shown in red (gray offline), and nodes in common to both networks are 
	considered to be coupled (shown by dashed lines). Highlighted in green 
	(gray offline), we represent a path between two nodes, the ``source'' $i$ 
	and the
	``sink'' $j$.}
	\label{fig:schematic}
\end{figure}

One broad class of processes that occur on a network are general
transport processes, or \textit{flows}~\cite{SC+08Oct}.  Whether flows
of people, fluids, or electrical currents, these systems can be
characterized by specifying the topology of the underlying network, a
source-sink distribution, and a dynamic (Fig.~\ref{fig:schematic}).
Where, to avoid confusion, we only imagine dynamical processes that
converge to a steady state---resulting in a stationary distribution of
flows over the network.  Unfortunately, the methods of analysis
mentioned above do not capture many of the typical features one might
expect here.  For example, it is easy to imagine a simple source-sink
distribution that allows the network to be split into two distinct
components such that the flows are unaffected.  In this case, the size
of the giant component may decrease but the network is still operating
well. With this example in mind, one question that arises is: how
should an interacting, or coupled, set of \textit{flow} networks be
characterized, and what are the interesting features of such systems?
Whilst any system-wide behaviour is intimately linked with the
particular dynamics, some understanding can be gained by investigating
the properties of simple examples that are chosen well enough to
represent some sub-class of these systems.  In this study, we report
the results of investigating such a toy model, and highlight the
interesting features which we believe might be typical of many
problems in this class.


\begin{figure}[h] \centering
	\subfigure[]{
	\label{subfig:road}
	\includegraphics[scale = 0.17]{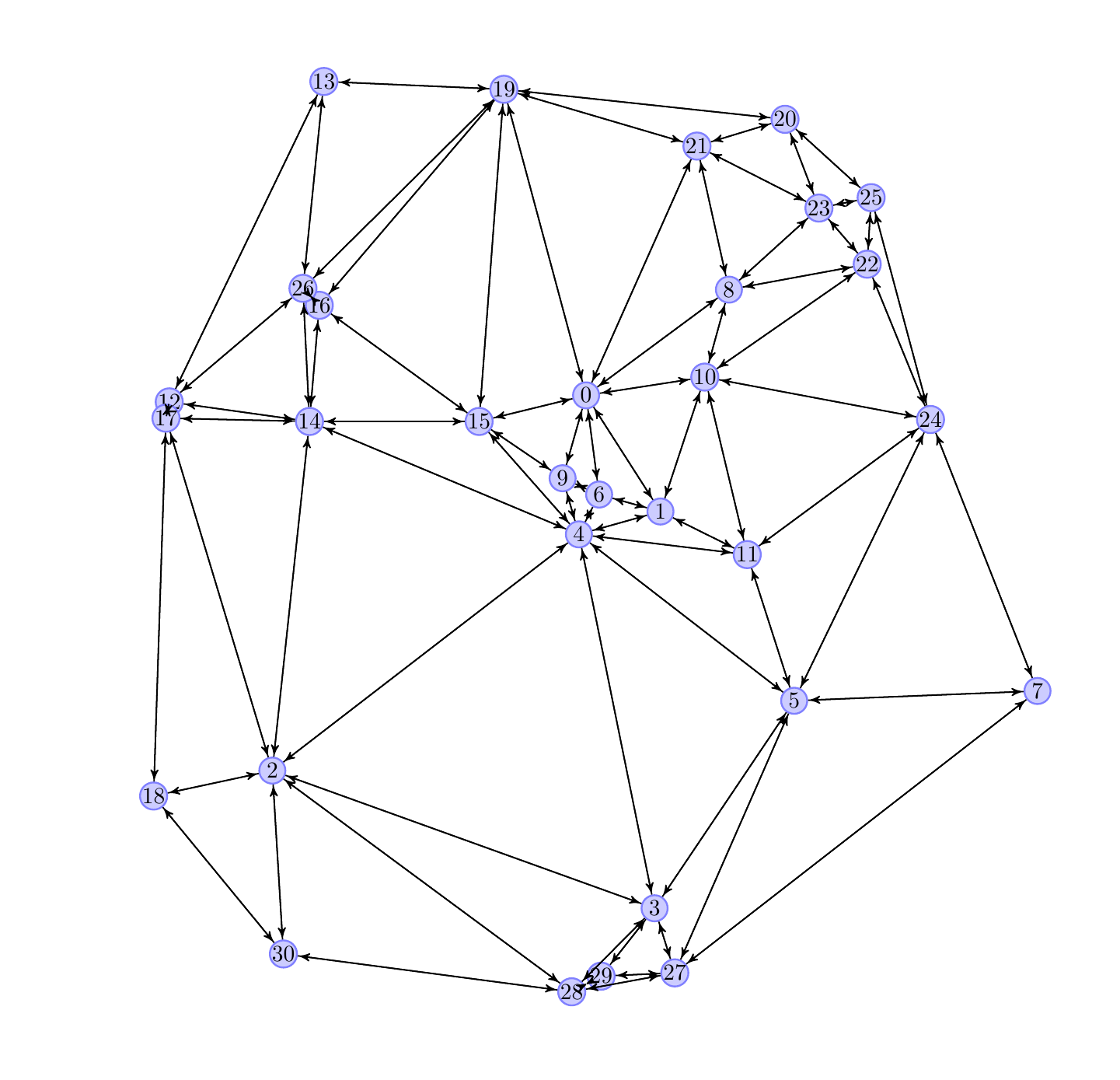}
	}%
	\subfigure[]{
	\label{subfig:rail}
	\includegraphics[scale = 0.17]{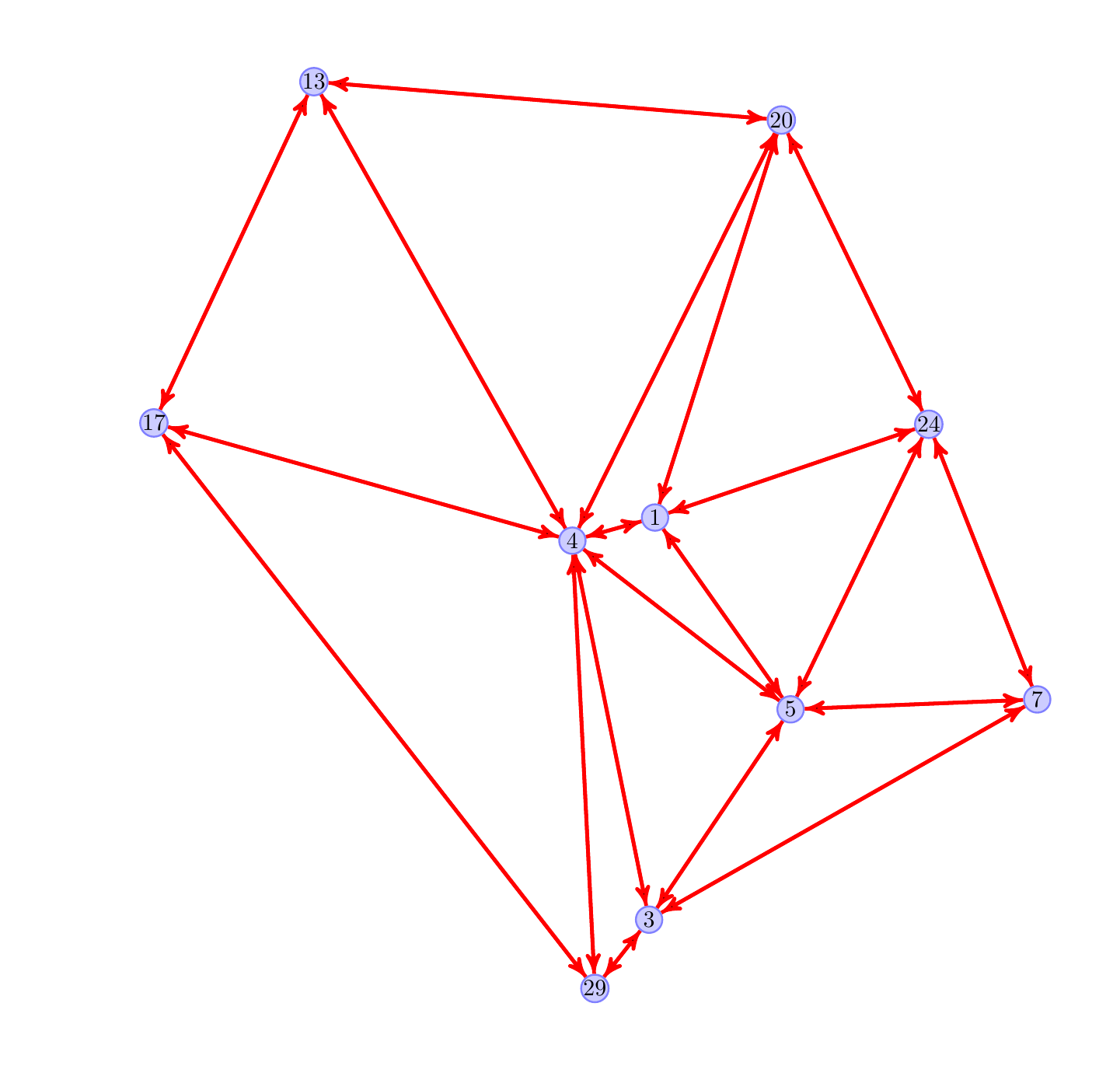}
	}
	\subfigure[]{
	\label{subfig:combined}
	\includegraphics[scale = 0.17]{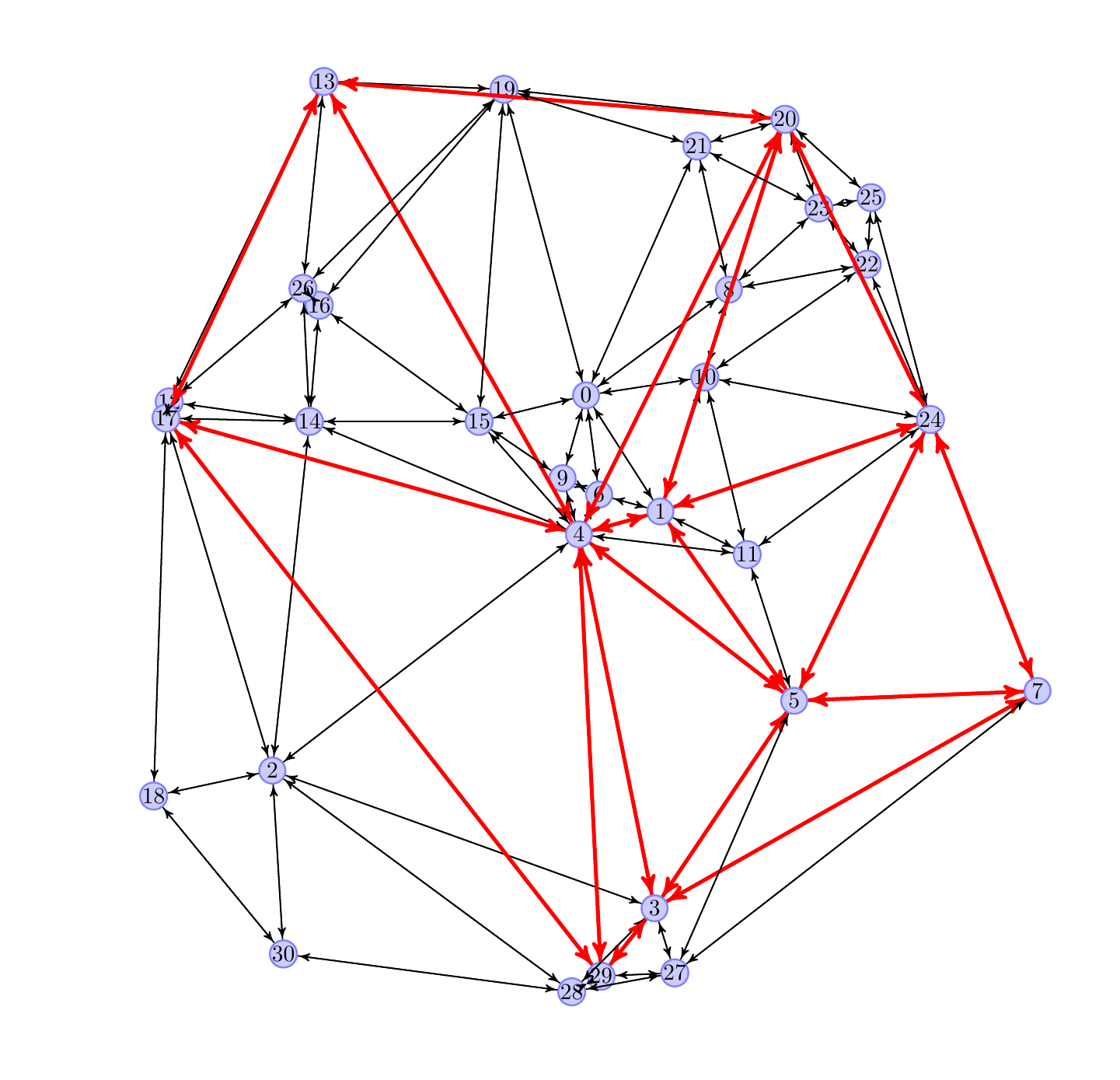}
	}
	\caption{(Color online) Each instance of the system is
          generated according to the following process:
		  \subref{subfig:road} First, $N_1$ nodes (here
		  $N_1=30$) are positioned at random within the unit circle and
          the Delaunay triangulation is produced; \subref{subfig:rail}
          the second network is then generated by drawing $N_2$ (here
          $N_2=10$) nodes uniformly from the existing ones ($N_2 \leq
		  N_1$) and, once again, computing the Delaunay triangulation;
		  \subref{subfig:combined} the combined system is no longer
		  planar and can be represented as a top-down view of 
		  Fig.~\ref{fig:schematic} (where zero weights assigned to the dotted 
		  interconnecting lines).
}
	\label{fig:example_network}
\end{figure}

Most existing studies of coupled networks focus on variants of the
random graph~\cite{SVB+10Apr,CDB+12Feb,CDB+12Apr}, primarily due to
the simplicity with which properties can be calculated.  However, many
physical networks (\textit{i.e}., electrical, transportation, ICT
\textit{etc}.)~are spatial networks and are often planar~\cite{MB11}.
For this reason, this letter focusses on coupled planar networks.
Although, for completeness, the example of spatially embedded
Erd\H{o}s-R\'{e}nyi random networks is discussed later alongside the
results for the planar case.  In addition to this, coupled networks
are usually found to be linked by a set of nodes common to both
networks (note that this is however not a necessary limitation of the
model, but simply a more realistic assumption for spatial networks).
For example, this is the case for a road network coupled to a rail or
subway network. Here, all the nodes of the road network are not nodes
of the rail network, but conversely, all stations are located at
points which can be considered as nodes in the road network. Motivated
by this simple example, we construct a first planar network as a
triangulation of points in the plane. Triangulations are often used as
a convenient way to generate planar networks from a given distribution
of nodes.  We choose the usual Delaunay triangulation~\cite{LGJS85},
which typically avoids slim triangles---not seen very often in real
networks due to their inefficiency---and which is effectively unique
for a given set of points. We then construct a second network based on
a random subset of the points used to construct the first network. Our
model thus comprises individual networks that are each planar Delaunay
triangulations, forming a combined network that is not necessarily
planar (see Fig.~\ref{fig:example_network}) and where the nodes of the
different networks with the same spatial location are linked together.
Rather than considering a dynamical system which acts to minimize a
global quantity---such as electrical networks, where the dissipated
power is minimized---we allocate flows on the network following a
basic transportation analogy.  Here, the source-sink distribution is
replaced by an origin-destination (OD) matrix $T_{ij}$.  This has the
benefit that it explicitly specifies the flow between node $i$ and
node $j$.  Therefore all that remains is to decide a method of route
assignment.  The obvious choice is to use the weighted shortest path,
where the number of such paths between nodes $i$ and $j$ is denoted by
$\sigma_{ij}$.  In our model, the weight associated with each edge
is the length of that edge multiplied by a factor $0 \leq \alpha_n
\leq 1$, which is common to all edges belonging to the same network.
The subscript $n$ is used to label the network: $n=1$ corresponds to
the larger network and $n=2$ the smaller.  The idea is that $\alpha =
\alpha_2 / \alpha_1$ is a single parameter that controls the relative
weight per unit distance between the two networks.  Indeed, in order
to simplify further, we impose the artificial constraint that
$\alpha\leq 1$.  This has the effect that a journey on the smaller
network ($n=2$) is favored over a journey of equivalent distance for
the larger network ($n=1$).  We also note that since the edge weights are 
proportional to edge length and nodes are positioned at random, it is very 
unlikely that $\sigma_{ij}>1$.

Previous studies of interacting networks use the term \textit{coupling} to describe how 
well two networks are linked.  Typically, this is a purely topological definition 
\textit{i.e}., the fraction of nodes from one network which link to 
another~\cite{RP+10Jul}, or the probability that a particular node has an edge which 
connects both networks~\cite{CDB+12Feb}.  For transport processes, a better measure of 
interaction must include details of how the flows are distributed.  For the system 
outlined above we define the coupling as
\begin{equation}
	\lambda \equiv \sum_{i \neq j} T_{ij} 
	\frac{\sigma_{ij}^{\mathrm{coupled}}}{\sigma_{ij}},
	\label{eq:lambda}
\end{equation}
where $\sigma_{ij}^{\mathrm{coupled}}$ is the number of shortest paths
between nodes $i$ and $j$, which include edges from both networks, and where 
the entries of the origin-destination matrix $T_{ij}$ are normalized 
\textit{i.e.}, $\sum_{ij} T_{ij} = 1$.  It is clear from
Eq.~(\ref{eq:lambda}) that $\lambda\in [0,1]$ is just the fraction of 
travellers that use both networks.  Such a definition is dependent on the 
method by which the flows are allocated and not just on the system topology.  
Indeed, for a given allocation method and network topology, there is usually a 
maximum value of $\lambda$ strictly less than one.  In our model, the coupling 
is controlled by choosing $\alpha$.  By virtue of changing the weights 
associated with each network, $\alpha$ changes the (weighted) shortest path 
between any two nodes.  For example, a value of $\alpha$ close to one indicates 
little difference between the two networks and hence, on average, shortest 
paths do not utilize both networks.  By contrast, a low value of $\alpha$ 
(close to zero) gives rise to significantly lower weights on the second network 
and therefore shortest paths typically use both networks.

With Eq.~(\ref{eq:lambda}) in mind, instead of investigating the likelihood
of catastrophic cascade failures, we consider more general measures of
how well the system is operating.  For example, one such measure is
the average distance travelled
\begin{equation}
	\bar{d} = \sum_{i\neq j} T_{ij} d_{ij},
	\label{eq:bard}
\end{equation}
where $d_{ij}$ is the distance travelled between nodes $i$ and $j$.
For most practical transport processes, a well designed system reduces
the average distance travelled (i.e., water/food supply, the Internet,
transportation, \textit{etc}.). Another important quantity, which is a
simple proxy for traffic, is the edge betweenness centrality, defined as
$x_e = \sum_{i\neq j} T_{ij} \left( \sigma_{ij}\left(e\right) /
  \sigma_{ij}\right)$,
where subscript $e$ is used to label edges, and $\sigma_{ij}\left(e\right)$ is the number 
of shortest paths between nodes $i$ and $j$, which use edge $e$. The
betweenness centrality allows us to introduce a second measure we are
concerned with, the Gini coefficient $G$.  A number between zero and one, $G$ is typically 
used in economics for the purpose of describing the distribution of wealth within a 
nation.  Here it is used to characterize the disparity in the assignment of flows to the 
edges of a network, something that has been done before for transportation systems such as 
the air traffic network~\cite{AR-F01}.  For example, if all flows were concentrated onto 
one edge, $G$ would be one, whilst if the flows were spread evenly across all edges, $G$ 
would be zero.  We use the definition according to Ref.~\cite{gini}
\begin{equation}
	G \equiv \frac{1}{2 E^2 \bar{x}} \sum_p \sum_q \vert x_p - x_q \vert,
	\label{eq:G}
\end{equation}
where subscripts $p$ and $q$ label edges, $E$ is the total number of edges, $x_p$ is the 
flow assigned to edge $p$ as defined earlier
, and $\bar{x} = \sum_p x_p / E$ is the average flow on an edge.

\begin{figure}[!h] \centering
	\subfigure[]{
	\label{subfig:od_mono}
	\includegraphics[scale = 0.17]{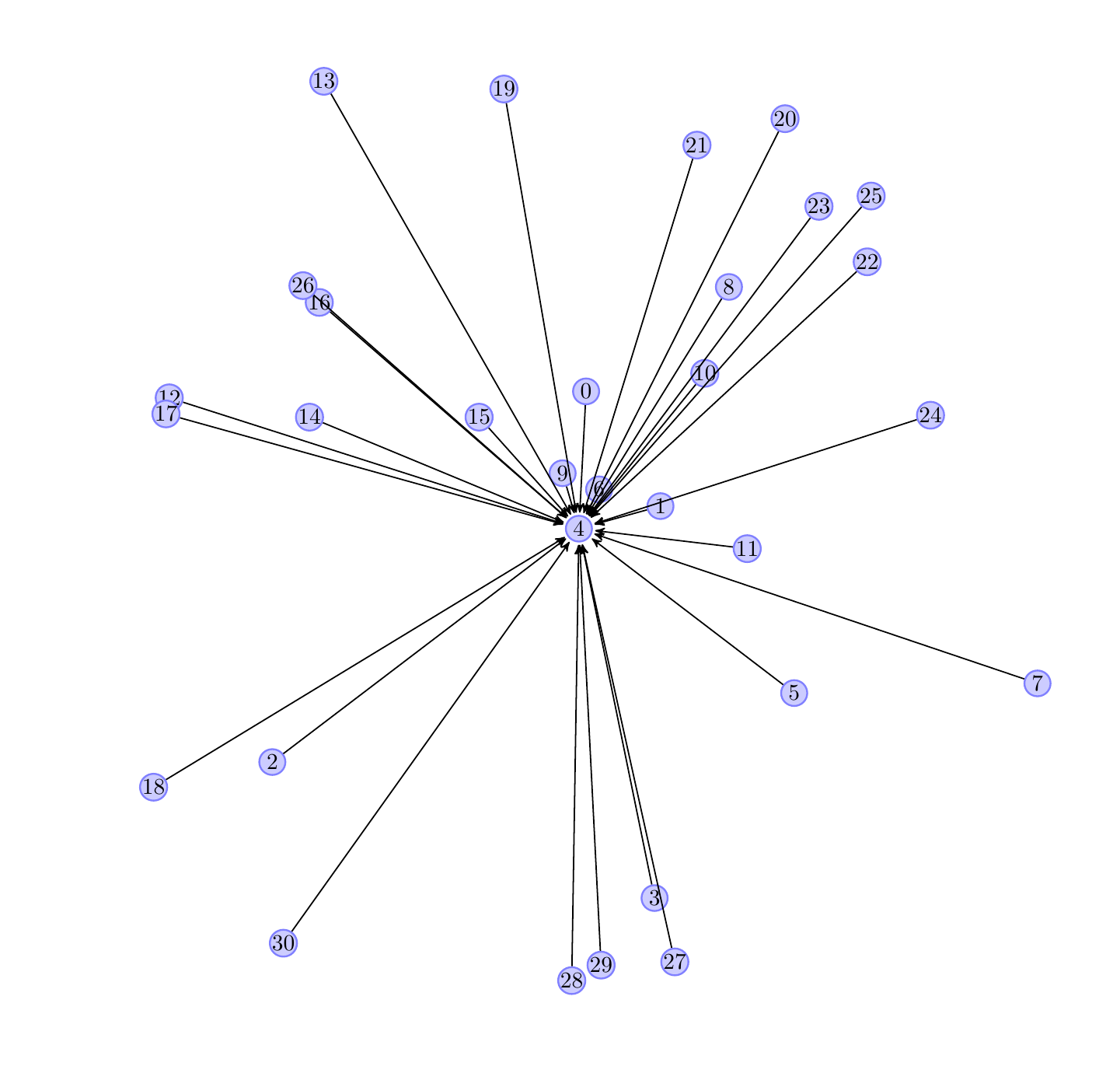}
	}%
	\hspace{8mm}
	\subfigure[]{
	\label{subfig:od_rewire}
	\includegraphics[scale = 0.17]{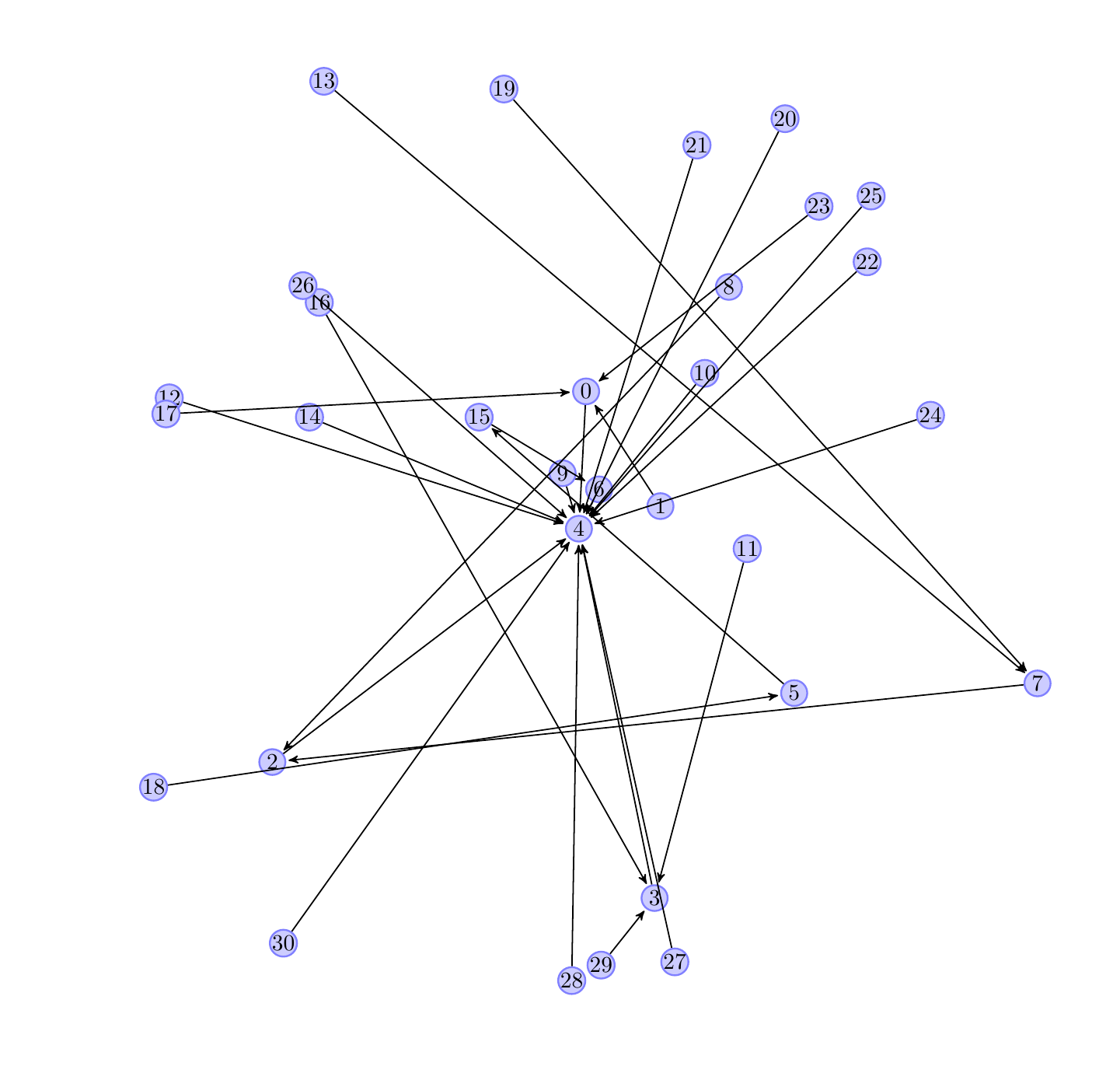}
	}%
	\caption{(Color online) Representations of OD matrices where
          each arrow corresponds to an entry in $T_{ij}$ and which
		  relates to the set of points in
          Fig.~\ref{fig:example_network}.  \subref{subfig:od_mono} A
          monocentric OD matrix. \subref{subfig:od_rewire} A
          monocentric OD matrix randomly rewired with probability
          $p=0.5$.}
	\label{fig:OD_matrices}
\end{figure}

Unfortunately, it is impractical to consider the interplay between $\lambda$, $\bar{d}$, 
and $G$, for all possible OD matrices.  Therefore it helps to choose a specific example.
We start with a monocentric OD matrix---\textit{i.e.}, all nodes travel to the 
origin---and then add noise by rewiring in the following way.  For each node, with 
probability $p$, choose a random destination, and with probability $1-p$, choose the 
origin (see Fig.~\ref{fig:OD_matrices}).

\begin{figure}[h] \centering
	\subfigure[]{
	\includegraphics[scale = 0.6]{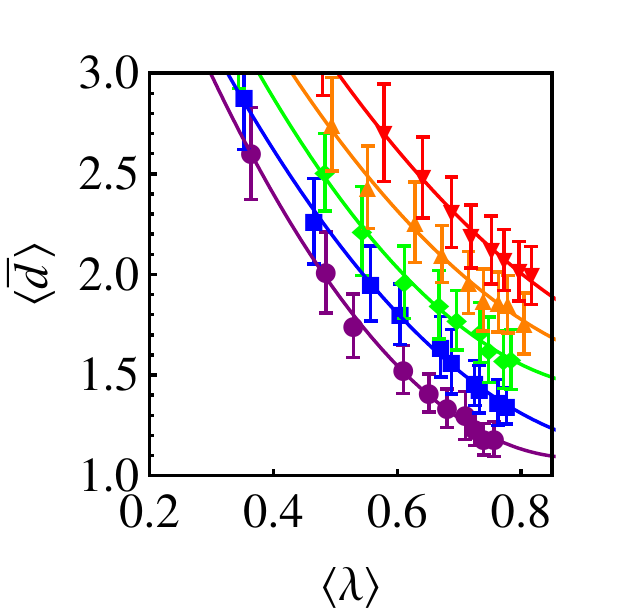}
	\label{subfig:bc_lam}
	}
	\subfigure[]{
	\includegraphics[scale = 0.61]{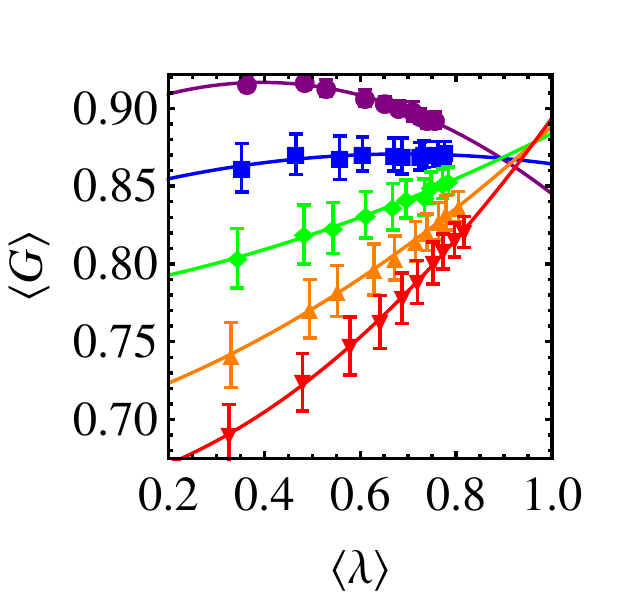}
	\label{subfig:gini_lam}
	}
	\caption{(Color online) Simulation results for the average
          shortest path and the Gini coefficient ($N_1 = 100$, $N_2 =
		  20$, and $p$ values: $0$ (purple dots), $0.2$ (blue squares), $0.4$
		  (green diamonds), $0.6$ (orange triangles), and $0.8$ (red inverted 
		  triangles)).  When the coupling
		  increases, the average shortest path decreases and the Gini
          coefficient can increase for large enough disorder.}
	\label{fig:bc_and_gin_lam}
\end{figure}

\begin{figure}[h] \centering
	\subfigure[]{
	\label{subfig:heatmap_p_low_w_high}
	\includegraphics[scale = 0.3]{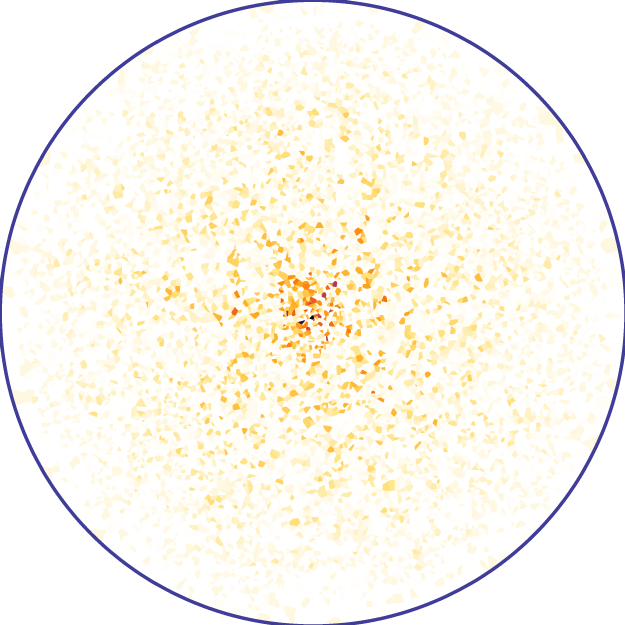}
	}%
	\subfigure[]{
	\label{subfig:heatmap_p_low_w_low}
	\includegraphics[scale = 0.3]{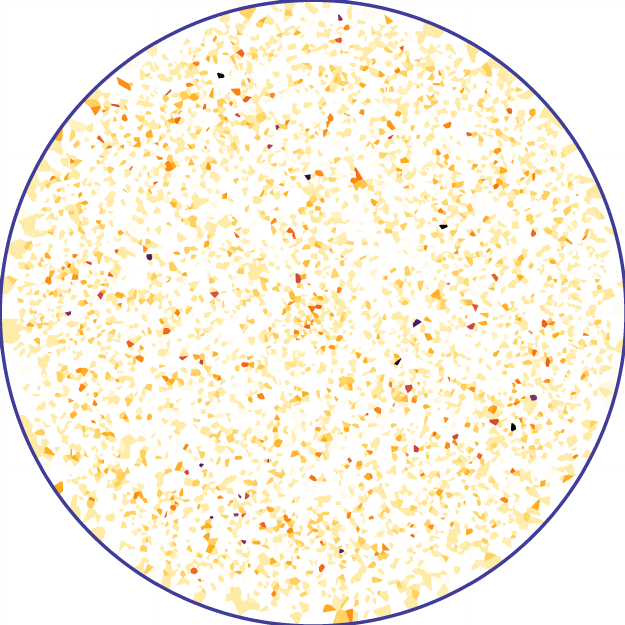}
	}
	\subfigure[]{
	\label{subfig:heatmap_p_high_w_high}
	\includegraphics[scale = 0.3]{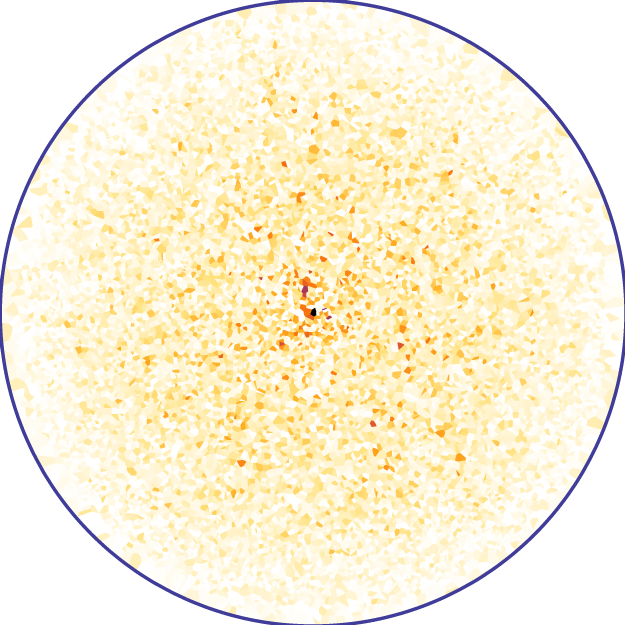}
	}%
	\subfigure[]{
	\label{subfig:heatmap_p_high_w_low}
	\includegraphics[scale = 0.3]{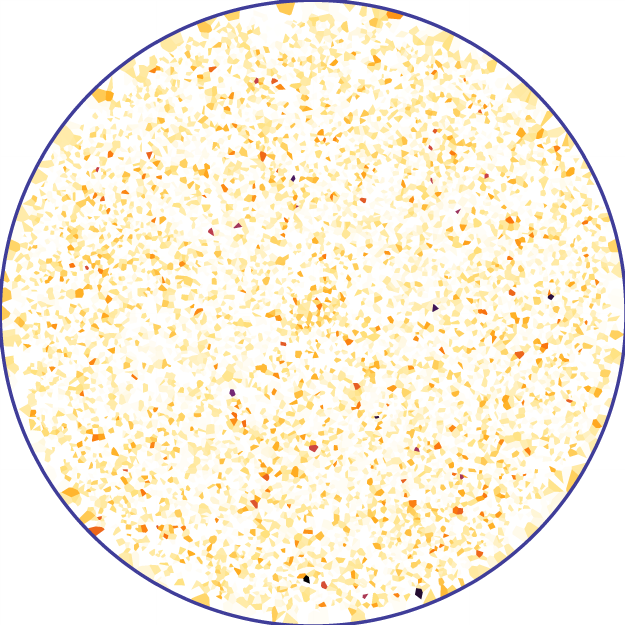}
	}%
	\caption{(Color online) Colormaps showing normalized edge flows---plotted at the 
	midpoint of each edge---over many instances of the system.  Colors are 
	assigned from lightest to darkest, starting with white (for zero flow) and 
	moving through yellow, orange and red for higher values of flow, until 
	reaching black (maximum flow).  Each Subfigure corresponds to the following 
	parameter values: \subref{subfig:heatmap_p_low_w_high} $p=0.2$, $\alpha = 
	0.9$; \subref{subfig:heatmap_p_low_w_low} $p=0.2$, $\alpha = 0.1$; 
	\subref{subfig:heatmap_p_high_w_high} $p=0.8$, $\alpha = 0.9$; 
	\subref{subfig:heatmap_p_high_w_low} $p=0.8$, $\alpha = 0.1$.}
	\label{fig:heatmaps}
\end{figure}

The set of numbers $N_1$, $N_2$, $p$, and $\alpha$, now define an ensemble of systems that 
are statistically equivalent (with respect to $\lambda$, $\bar{d}$, and $G$).  We proceed 
by calculating the quantities $\langle \lambda\rangle$, $\langle \bar{d}\rangle$, and 
$\langle G\rangle$ for different values of $p$ and $\alpha$, where angle brackets 
$\langle\dots\rangle$ represent an ensemble average.  The results are shown in 
Fig.~\ref{fig:bc_and_gin_lam}, where each data point corresponds to an average over fifty 
instances of the OD matrix for each of fifty instances of the coupled network geometry.
We find that, as the coupling increases, the length of the average
shortest path decreases (Fig.~\ref{subfig:bc_lam}).  This is
straightforward to understand since the increased coupling is simply a
result of reducing $\alpha$.  Furthermore it is clear that increasing
randomness in the origin destination matrix increases the length of
the average shortest path by an almost constant value, irrespective of
the coupling. By contrast, the behaviour of the Gini coefficient at
different couplings (Fig.~\ref{subfig:gini_lam}) is less easily
explained.  Consider instead Fig.~\ref{fig:heatmaps}.  Here, each
colormap shows the distribution of flows resulting from many instances
of the system.

The first two plots, Figs.~\ref{subfig:heatmap_p_low_w_high} and
\ref{subfig:heatmap_p_low_w_low}, were generated from OD matrices
rewired with low probability ($p=0.2$) \textit{i.e.}, almost
monocentric.  The ratios of edge weights per unit distance between the
two networks are $\alpha=0.9$ and $\alpha=0.1$ respectively.
Therefore each diagram corresponds to a point on the blue line in
Fig.~\ref{subfig:gini_lam}.  For $\alpha=0.9$, there is minimal
coupling between the networks and a high concentration of flows are
seen around the origin.  Since the flows are disproportionately
clustered, this configuration is described by a high Gini coefficient.
By contrast, for $\alpha=0.1$, the difference in the edge weights
means that it can be beneficial to first move away from the origin in
order to switch to the `fast' (low $\alpha$) network.  We therefore see a broader distribution of
flows with small areas of high concentration around coupled nodes.
The emergence of these \textit{hotspots} away from the center also
corresponds to a high Gini coefficient---and therefore the blue line
in Fig.~\ref{subfig:gini_lam} is relatively
flat. Figs.~\ref{subfig:heatmap_p_high_w_high} and
\ref{subfig:heatmap_p_high_w_low} correspond to the red line of
Fig.~\ref{subfig:gini_lam}: generated from OD matrices rewired with
high probability ($p=0.8$).  We observe that even for $\alpha$ close
to one, the localization of flows is less than for
$p=0.2$---resulting in a lower Gini coefficient.  As $\alpha$ is
decreased, the second network becomes more favourable and coupling
\textit{hotspots} can be seen once again---resulting in a high Gini
coefficient and a positive gradient for the red line of
Fig.~\ref{subfig:gini_lam}. This result points to the general idea
that randomness in the source-sink distribution leads to local
congestion and more generally to a higher sensitivity to coupling.

\begin{figure}[h] \centering
	\subfigure[]{
	\includegraphics[scale = 0.62]{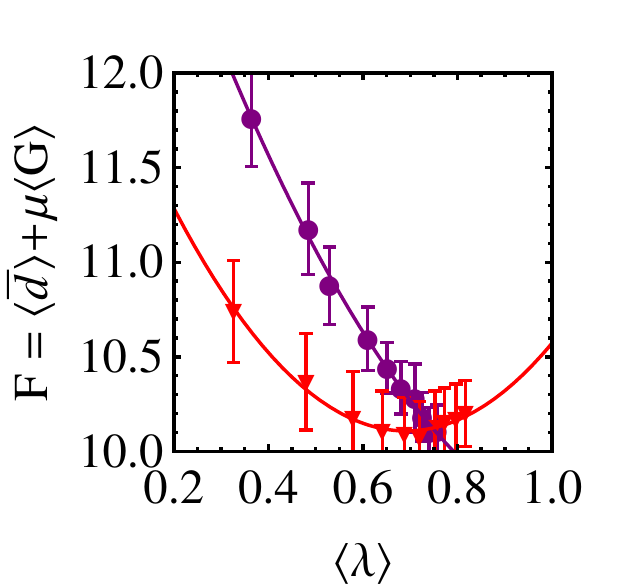}
	\label{subfig:F_lam}
	}
	\subfigure[]{
	\includegraphics[scale = 0.6]{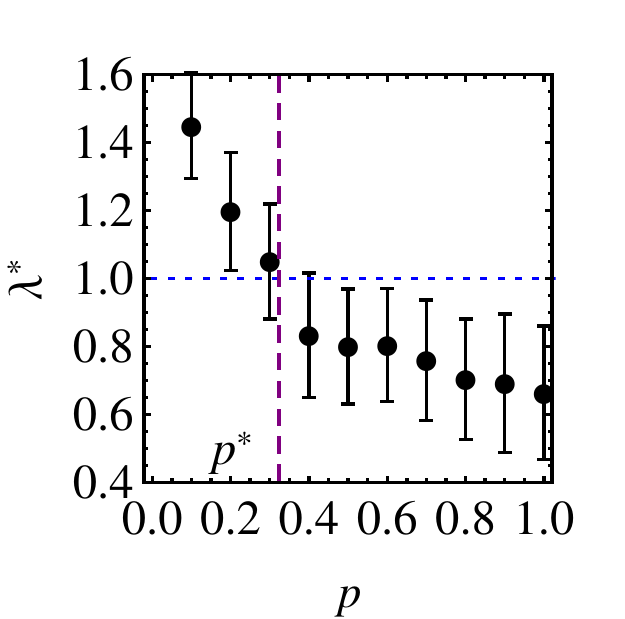}
	\label{subfig:lamstar}
	}
	\caption{Existence of an optimal coupling:
          \subref{subfig:F_lam} Simulation results for $\mu = 10$,
		  $N_1 = 100$, $N_2 = 20$, and $p$ values: $0$ (purple dots), and $0.8$ 
		  (red inverted triangles) (only two values of $p$ are shown
		  to ensure the lines of best-fit can be seen
          clearly). \subref{subfig:lamstar} Minima of quadratic
		  best-fit curves for different values of
		  $p$. We obtain $p^* \simeq 0.32$ using a straight-line approximation 
		  between the two closest points, above and below, $\lambda^*=1$.  (The 
		  error
          bars shown are those of the closest data point to the
          minimum of the best-fit curve).}
	\label{fig:F_lam_and_lamstar}
\end{figure}

Heuristically, one might consider this a simple model of a two mode
transportation system in the low density regime---\textit{i.e.}, where
the effects of congestion do not affect route choice.  We imagine a
road network coupled to a rail network where users select the quickest
route to their destination.  That is, the shortest path actually
represents the quickest path.  This implies that the scale factors
$\alpha_\mathrm{road}$ and $\alpha_\mathrm{rail}$ must have units of
time divided by distance, so we assume that they represent the inverse
of the average speed associated with each mode.  The result is that
decreasing (increasing) the ratio of these factors, $\alpha$, has the
effect of increasing (decreasing) the relative speed of rail above
road---and hence the coupling. In this picture, the Gini coefficient
can now be thought of as a measure of road use.  A low value indicates
that the system uses all roads to a similar extent, whilst a high
value indicates that only a handful of roads carry all the traffic.
With this analogy in mind, it is natural to combine the effects
observed above into a single measure.  We assert that it is likely that a
designer or administrator of a real system would wish to
simultaneously reduce the average travel time and minimize the
disparity in road utilization. To serve this purpose, we define a 
`utility' function $F = \langle\bar{d}\rangle +
\mu\langle G \rangle$, where it is immediately apparent from
Fig.~\ref{fig:bc_and_gin_lam} that, for certain values of $\mu$, the
function $F$ will have a minimum.  That is, a non-trivial
(\textit{i.e.}, non-maximal) optimum $\lambda$ will emerge.  Figure
\ref{subfig:F_lam} shows that, whether a non-trivial optimum coupling
exists depends on the origin-destination matrix.  For OD matrices
rewired with a high probability, increasing the speed of the rail
network reduces the road utilization as flows become concentrated
around nodes where it is possible to change modes.  Dependent on the
value of $\mu$, the effect of reduced utilization can outweigh the
increased journey time, leading to a minimum in $F$.  Monocentric OD
matrices, by contrast, have inherently inefficient road utilization
when applied to planar triangulations, regardless of the speed of the
rail network.  Therefore no minimum is observed, and hence no
(non-trivial) optimum $\lambda$.
More systematically, we plot in Fig.~\ref{subfig:lamstar} the
minima $\lambda^*(p)$ of quadratic best-fit curves obtained from
Fig.~\ref{subfig:F_lam}, each corresponding to a different value of $p$.
Defining $p^*$, the value of $p$ for which $\lambda^*(p^*) = 1$, it is
possible to categorize the system into one of two regimes.  We observe
that: if $p<p^*$, then the optimal coupling is trivially the maximum;
otherwise if $p\geq p^*$, a non-trivial optimal coupling exists.

Finally, we note that similar effects can be observed on other
non-planar networks.  As mentioned earlier, one example is
Erd\H{o}s-R\'{e}nyi random networks generated for a set of nodes with
random locations.  Whilst the observed behaviour is qualitatively the
same, the results are much less pronounced due to the absence of
spatial structure and the spatial localization of centrality.

In conclusion, the model is characterised by two competing
\textit{forces}---the desire to move all flows onto the most efficient
network, whilst also ensuring that congestion does not arise around
the nodes which connect both networks.  We observe that the
optimisation of such a system can be sensitive to randomness
introduced in the origin-destination matrix.
The broader interpretation of our work is that, spatial, space-filling, 
networks such as
transportation networks or the electricity grid, may be inherently
fragile to certain changes in supply and demand, such as the
transition from centralized power generation to decentralized
\textit{prosumers}~\cite{IL+10}.  This behaviour is not captured by
the existing literature, and demonstrates an alternative view of
transport processes on interacting networks.  Indeed, whilst the
assumptions made have a convenient interpretation in terms of bimodal
transportation systems, we expect that the results hold for a broader
class of systems and welcome work in this area.

Acknowledgements. 
RGM thanks the grant CEA$/$DSM-Energie for financial support.

\bibliographystyle{prsty}

\end{document}